# On the role of speed adaptation and spacing indifference in traffic instability: evidence from car-following experiments and its stochastic modeling

Junfang Tian[1], H.M.Zhang[2,*], Martin Treiber[3], Rui Jiang[4,*], Zi-You Gao[4], Bin Jia[4]

[1]Institute of Systems Engineering, College of Management and Economics, Tianjin University
No. 92 Weijin Road, Nankai District, Tianjin 300072, China

[2] Department of Civil and Environmental Engineering, University of California Davis, Davis, CA 95616, United States

[3]Technische Universität Dresden, Institute for Transport & Economics, Würzburger Str. 35, D-01062 Dresden, Germany

[4]MOE Key Laboratory for Urban Transportation Complex Systems Theory and Technology, Beijing Jiaotong University
No.3 Shangyuancun, Haidian District, Beijing 100044, China

Understanding the mechanisms responsible for the emergence and evolution of oscillations in traffic flow has been subject to intensive research by the traffic flow theory community. In our previous work, we proposed a new mechanism to explain the generation of traffic oscillations: traffic instability caused by the competition between speed adaptation and the cumulative effect of stochastic factors. In this paper, by conducting a closer examination of car following data obtained in a 25-car platoon experiment, we discovered that the speed difference plays a more important role on car-following dynamics than the spacing, and when its amplitude is small, the growth of oscillations is mainly determined by the stochastic factors that follow the mean reversion process; when its amplitude increases, the growth of the oscillations is determined by the competition between the stochastic factors and the speed difference. An explanation is then provided, based on the above findings, to why the speed variance in the oscillatory traffic grows in a concave way along the platoon. Finally, we proposed a mode-switching stochastic car-following model that incorporates the speed adaptation and spacing indifference behaviors of drivers, which captures the observed characteristics of oscillation and discharge rate. Sensitivity analysis shows that reaction delay only has slight effect but indifference region boundary has significant on oscillation growth rate and discharge rate.

[*] Corresponding authors.

Email address: hmzhang@ucdavis.edu, jiangrui@bjtu.edu.cn

## 1. Introduction

The emergence and evolution of traffic flow oscillations as well as their development into jams is an intriguing pattern formation phenomenon (e.g. Schönhof and Helbing, D. 2007; Kerner 2004, 2009) that has undesirable consequences, because the resulting stop-and-go movement is a nuisance to motorists, consumes more fuel, and likely causes more accidents. The true origin of these traffic waves, however, is still under debate. To explain the causes of the formation and development of traffic oscillations as well as other traffic phenomena, several traffic models have been proposed, based on which mathematical analyses have been performed. These models can be roughly divided into two categories: macroscopic ones (e.g. Payne, 1971; Kerner and Konhäuser, 1993; Jiang et al., 2002) and microscopic ones (e.g. Bando et al., 1995; Kerner and Klenov, 2002; Jiang et al. 2001; Li and Sun, 2012; Chen et al., 2012; Tian et al., 2016a). The macroscopic models arise based on hydrodynamic method and are formulated by partial differential equation or equations. The microscopic models aim to describe driver's driving behavior, based on the driver's perception to the nearby driving conditions.

A Car Following Model (CFM) is an important type of microscopic models, which describes the longitudinal interaction between vehicles, usually based on ordinary differential equations. One of the earliest CFM is the General Motor (GM) model (Chandler et al., 1958), which uses the speed difference $\Delta v$ between the current vehicle and the preceding vehicle to calculate the delayed acceleration. The simplest GM model is as follows

$$a(t)=\lambda \Delta v(t-\tau) \tag{1}$$

where $\lambda$ is sensitivity parameter and $\tau$ is time delay. In the steady state of this model, the speed of vehicles equals to each other, and traffic flow is unstable when $\lambda\tau>0.5$. In this case, disturbance will grow over time along a platoon of vehicles.

Later many CFMs were proposed, among which the Gipps model (Gipps, 1981), the optimal velocity (OV) model (Bando, 1995), the full velocity difference (FVD) model (Jiang, 2001), and the intelligent driver model (Treiber et al., 2000) are some representative ones. In the steady state of these models, the speed and spacing of vehicles equal to each other, and there is a unique relationship between the speed and spacing. For example, the FVD model can be regarded as a combination of the OV model and the GM model, which is as follows

$$a(t)=\kappa\left(V_{op}(d)-v\right)+\lambda\Delta v(t) \quad (2)$$

where $V_{op}(d)$ is the optimal velocity determined by the spacing $d$. In the special case $\lambda=0$, the FVD model reduces to the OV model. In the unstable situation $V'_{op}(d)<\kappa/2+\lambda$, a disturbance will gradually grow and develop into low speed vehicle cluster or clusters. The instability of such models has been believed to be the formation mechanism of traffic jams for more than half a century. In these models, it has been proven that the oscillations grow initially in a convex way, see Li et al. (2014).

The high-order macroscopic models have similar instability mechanism to the car-following models. For example, the speed gradient model (Jiang et al., 2002) is as follows

$$\frac{\partial \rho}{\partial t}+\frac{\partial(\rho u)}{\partial x}=0 \quad (3)$$

$$\frac{\partial u}{\partial t}+u\frac{\partial u}{\partial x}=c_0\frac{\partial u}{\partial x}+\frac{u_e-u}{\tau} \quad (4)$$

where $\rho=\rho(x,t)$ and $u=u(x,t)$ are the density and speed at position $x$ and time $t$. $c_0$ is the propagation speed of small disturbance, $\tau$ is relaxation time, $u_e=u_e(\rho)$ is the equilibrium speed-density relationship. The relation $u_e(\rho)$ determines the steady state and unique relationship between speed and density. The traffic flow is unstable when $|\rho u'_e(\rho)|\leq c_0$.

On the other hand, cellular automaton (CA) models (e.g. Nagel and Schreckenberg, 1992; Tian et al., 2015, 2016b, 2017) are a different type of microscopic model from the CFM. In the CA models, the time, the space and the state are all discrete. One main feature of CA models is the introduction of randomization, which enables the models to depict the spontaneous formation of jams. In the CA models, the formation and growth of oscillations is due to stochastic factors, which is different from the traditional CFM and high-order macroscopic models.

Recently, a series of car-following experiment were performed on open roads (Jiang et al., 2014, 2015, 2018). Stripe structures have been observed in the spatiotemporal evolution of traffic flow, which corresponds to the formation and development of oscillations behind a "moving bottleneck". It has been found that the standard deviations of speed increase in a concave way along the platoon. Later, this concave growth pattern of oscillations has also been found in traffic on the US 101 highway (Tian et al., 2016c) and in US experiment (Zheng et al., 2018). Therefore, the formation mechanism of traffic oscillations in traditional models was under question.

Based on the experimental findings, Jiang et al. (2014) proposed a 2-Dimensional Intelligent Driver Model (2D-IDM), in which the desired time gap changes stochastically with time. As a result, there is no steady state and the traffic state of a following vehicle could span a 2D region in the speed-spacing plane, even if the leading vehicle moves with constant speed. 2D-IDM can reproduce the concave growth pattern of traffic oscillations and the simulation results fit the experimental ones quantitatively well. Therefore, Jiang et al. (2018) proposed a new instability mechanism, i.e., traffic instability is determined by competition between the speed adaptation effect and the cumulative effect of stochastic factors.

In this paper, we make a thorough analysis of the trajectories of the car-following experimental data, which supports the proposed instability mechanism. It reveals that the time series of acceleration and the speed difference exhibit striking similarity, and the speed difference plays a more important role on car-following dynamics than spacing. Moreover, it is shown that the stochasticity follows a mean reversion process. Based on the findings, we propose a heuristic analysis to explain why oscillation grows in a concave pattern. Finally, we propose a new hybrid, stochastic car-following model. Simulation results show that speed adaptation and stochasticity lead to oscillation formation.

The paper is organized as follows. Section 2 performs the data analysis. Section 3 presents the heuristic analysis. Section 4 proposes the new car-following model. Section 5 shows the simulation results. Section 6 concludes the paper.

## 2. Experimental data analysis

In this paper, the stationary states of the 25-car-platoon experiments (Jiang et al., 2014, 2015) will be extracted and analyzed. Since the GPS devices record the speed every $dt = 0.1$ s, we calculate the acceleration via

$$a_n(t) = \frac{v_n(t) - v_n(t - dt)}{dt} \tag{5}$$

where $a_n(t)$ and $v_n(t)$ are, respectively, the acceleration and speed of vehicle $n$ at time $t$. To reduce random fluctuations, we employ the moving average method with a 1-second time window. It should be mentioned that when the time window reaches 2 s, the true acceleration peaks have been damped. We have examined the results by changing the time windows in the range between 0.5 s and 2 s, and found only minor quantitative differences in the results. Thus, we choose a time window of 1 s to the smooth the acceleration rates.

Then we compare the time series of acceleration with the speed difference between a car and its front car. Fig.1 shows two typical examples. One can see that the two time series exhibit striking similarity, and a time delay of acceleration time series is manifested. To extract the time delay, the correlation method is used

$$\tau_n = \underset{\tau' \in (0,5)}{\arg\max} \left( \mathrm{corr}\left( a_n^{\exp}(t), \Delta v_n^{\exp}(t - \tau') \right) \right) \tag{6}$$

Fig.2(a) shows the statistical results of extracted time delay of the 24 cars. Fig.2(b) shows the distribution of the mean time delay of each driver. The Jarque-Bera test shows that it follows the lognormal distribution ($p = 0.213$). Fig.3 shows $R^2$ between $\Delta v_n^{\exp}(t - \tau_n)$ and $a_n(t)$. As a comparison, we also calculated $R^2$ between $d_n^{\exp}(t - \tau_n)$ and $a_n^{\exp}(t)$. One can see that $R^2$ is much larger between $\Delta v_n^{\exp}(t - \tau_n)$ and $a_n(t)$, which indicates that the speed difference plays a more important role on car-following dynamics, at least in the spacing range concerned, see Fig.4.

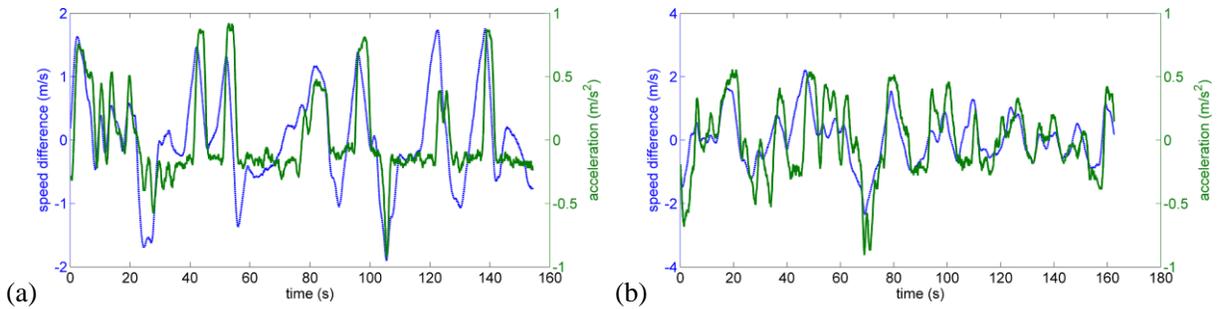

**Figure 1.** Two samples of the speed difference and acceleration time series.

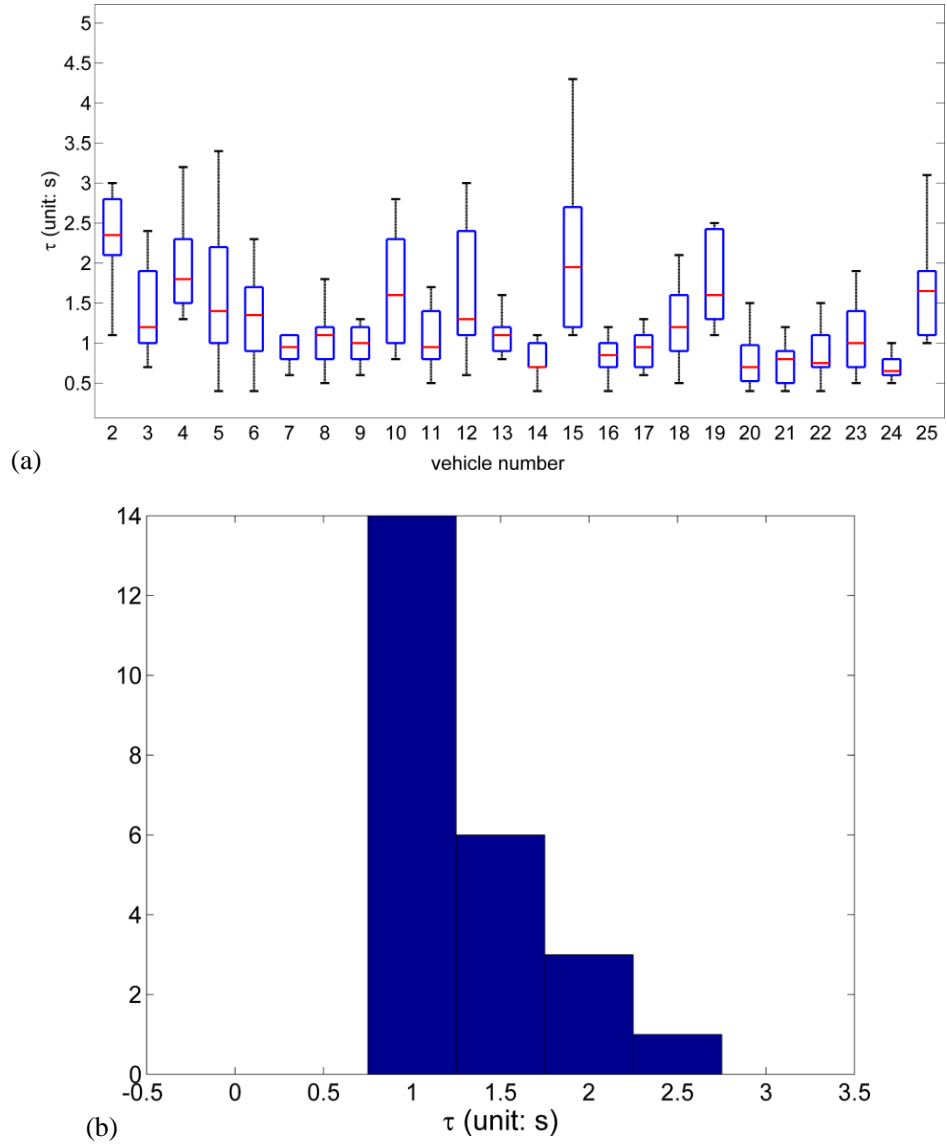

**Figure 2.** The time delay $\tau$ of the 24 cars. (a) distribution of time delay of each car in each run; (b) distribution of mean time delay of the cars over all runs.

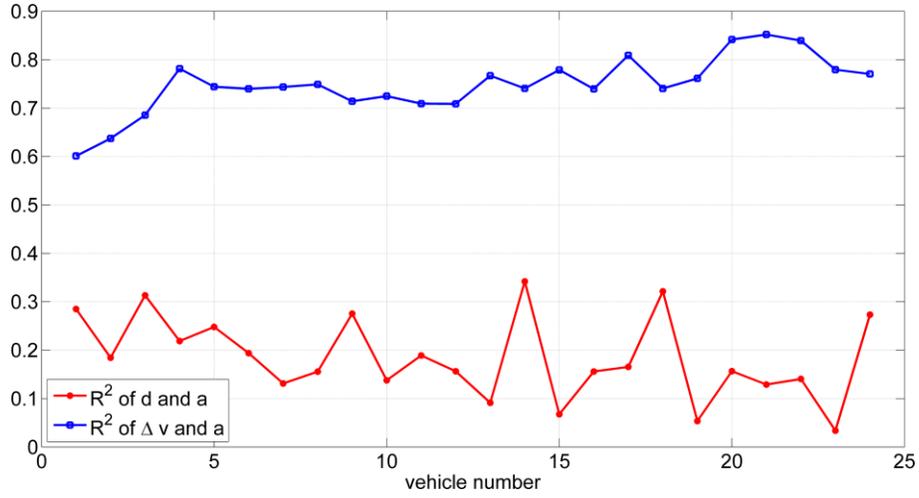

**Figure 3.** $R^2$ between speed difference $\Delta v$ and acceleration and $R^2$ between spacing $d$ and acceleration.

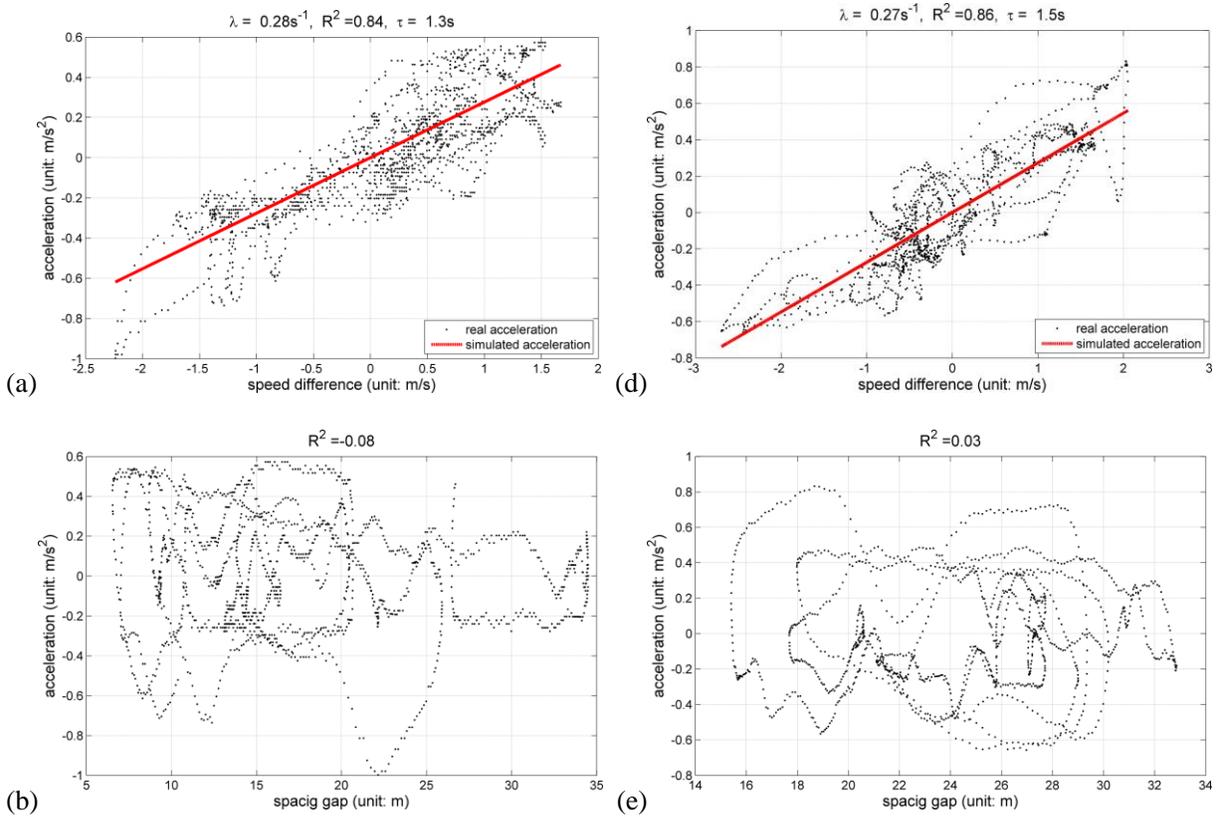

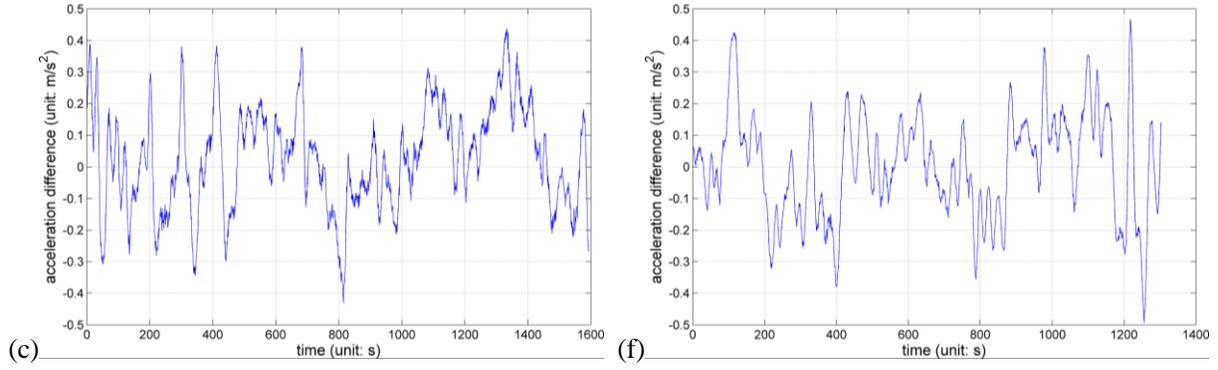

**Figure 4.** Two samples of (a, d) acceleration $a(t)$ vs. speed difference $\Delta v(t\text{-}\tau)$, (b, e) acceleration $a(t)$ vs. spacing $d(t\text{-}\tau)$, and (c, f) the time series of $\xi(t)$. In (a, d) the red line denotes $a = \lambda \Delta v(t\text{-}\tau)$.

Next, $\lambda_n$ is calculated through linear regression of $a_n(t) = \lambda_n \Delta v_n(t - \tau_n)$, see Fig.4 (a, d) for two examples. As a comparison, we show acceleration vs spacing in Fig.4 (b, e), which again indicates that the dependence of acceleration on spacing is not so remarkable as on speed difference. The statistical results of $\lambda_n$ are shown in Fig.5(a). Fig. 5(b) shows the distribution of mean sensitivity of each driver. The Jarque-Bera test shows that it also follows the lognormal distribution ($p = 0.3729$).

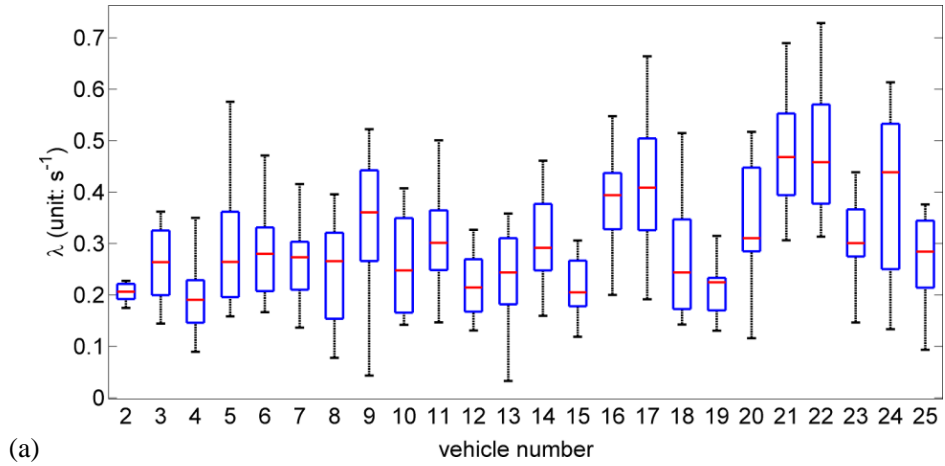

(a)

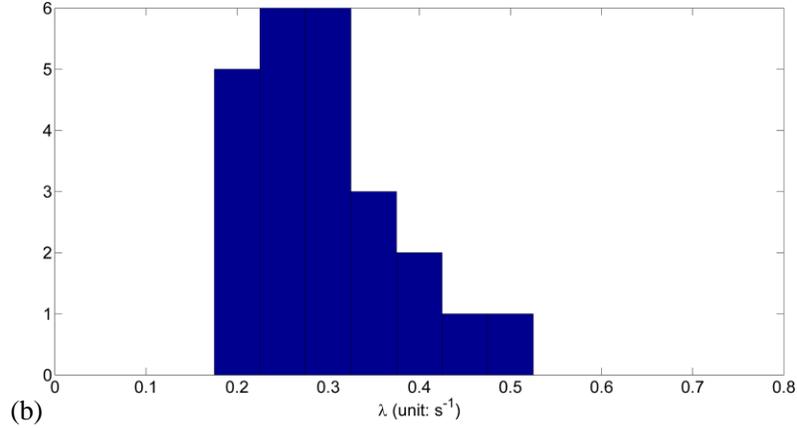
(b)

**Figure 5.** The distribution of sensitivity $\lambda$. (a) distribution of sensitivity of each car in each run; (b) distribution of mean sensitivity of the cars over all runs.

Now we subtract $\lambda\Delta v(t-\tau)$ from the acceleration. Denote the resulted residual time series as $\xi(t)$. Fig.4 (c, f) shows two typical examples. We performed an augmented Dickey–Fuller test (ADF) test of $\xi(t)$. The result is shown in Fig.6. One can see that in most cases, $\xi(t)$ is stationary, therefore $\xi(t)$ can be regarded as a Mean Reversion process (Balvers, et al. 2000).

To measure the strength of $\xi(t)$, the standard deviations of $a(t)$ and $\xi(t)$ are calculated. Fig.7(a) shows that both $\sigma_a$ and $\sigma_\xi$ grow concavely along the platoon. Fig.7(b) show that (1) the ratio $\sigma_\xi/\sigma_a$ is quite large for the several vehicles in the front part of the platoon, which indicates that when its amplitude is small, the growth of the oscillations is mainly determined by the stochastic factors, since the speed difference among vehicles are small in the front part of the platoon; (2) this ratio decreases along the platoon, since the speed difference grows along the platoon. Therefore, the growth of the oscillations is determined by the competition between the stochastic factors and the speed difference.

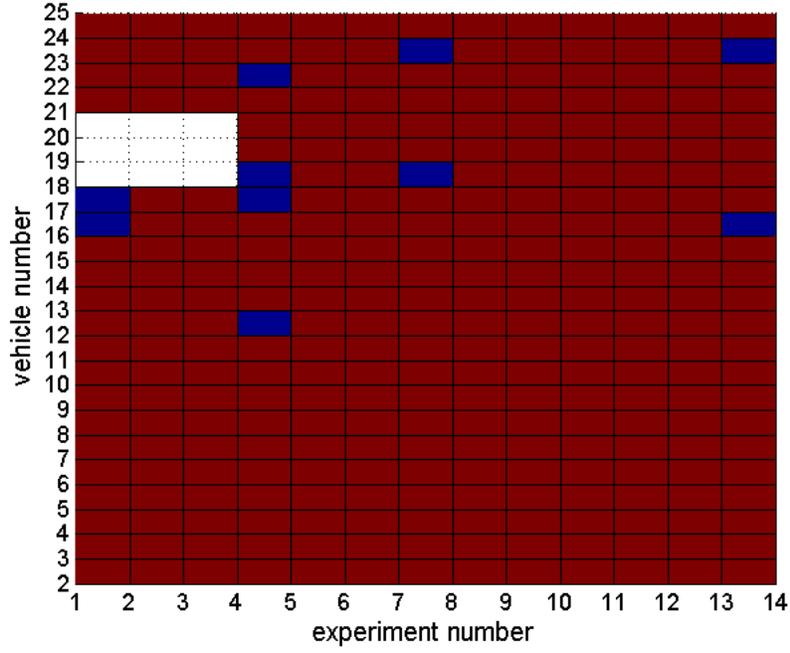

**Figure 6.** The result of the augmented Dickey–Fuller (ADF) test. The red grid means that $\xi(t)$ is stationary while the blue means nonstationary. The blanks are the missing data.

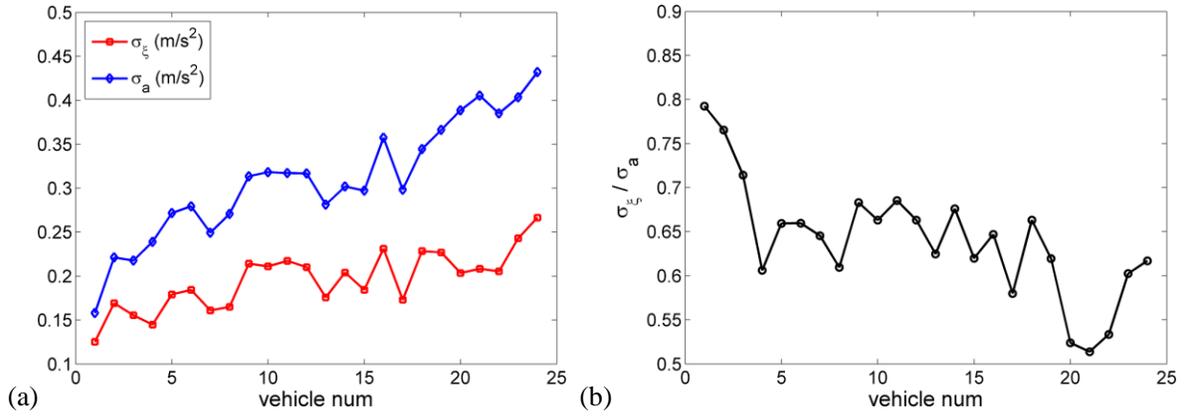

**Figure 7.** (a) The standard deviation of the acceleration $\sigma_a$ and residual $\sigma_\xi$; (b) the ratio of $\sigma_\xi/\sigma_a$.

Therefore, we use Vasicek model (Vasicek, 1977) to approximate $\xi_n(t)$. This model is the first one to capture mean reversion process and originally used to describe the evolution of interest rate derivatives, which reads

$$d\xi_n = \kappa\left(\mu - \xi_n\right)dt + \sigma d\mathrm{W}_n \tag{7}$$

where $W_n(t)$ is a Wiener process, $dW_n$ follows the normal distribution with mean zero and variance d$t$. $\sigma$ denoted the instantaneous volatility measuring the amplitude of randomness. $\mu$ is the long term mean level, $\xi_n(t)$ will evolve around the mean level $\mu$ in the long run. $\kappa$ is the speed of reversion, which characterizes the speed at which $\xi_n(t)$ returns to $\mu$. It is worth to mention that $\sigma^2/2\kappa$ is the long term variance. $\xi_n(t)$ will regroup around $\mu$ with this variance after a long time.

To calibrate $\kappa$ and $\sigma$, the model is reformulated into the discrete form

$$\xi_n(t) = (1-\kappa dt)\xi_n(t-dt) + \kappa\mu dt + \sigma dW_n \tag{8}$$

The least square method is used in the calibration. The statistical results are shown in Fig.8. The mean values of $\kappa$ and $\sigma$ are 0.11 and 0.08, respectively, over all drivers. Fig.8(c) shows that the range of the mean $\mu$ is quite small and very close to zero. This is reasonable, and one can expect that $\mu$ tends to zero if the time series becomes longer. Thus, the acceleration residual $\xi_n(t)$ can be approximated by

$$d\xi_n = -\kappa\xi_n dt + \sigma dW_n \tag{9}$$

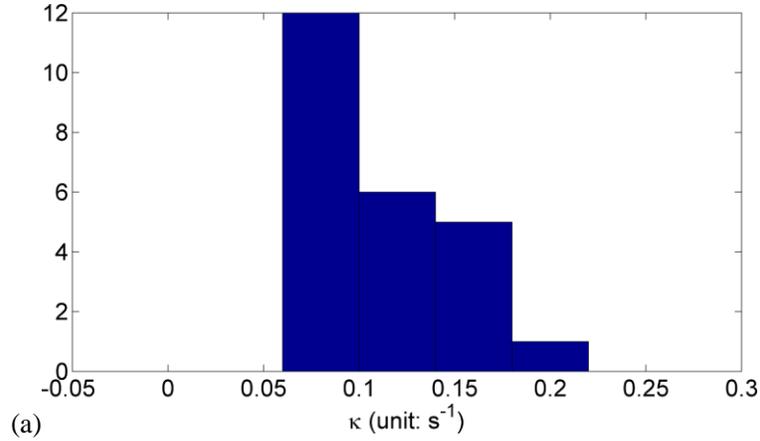

(a)

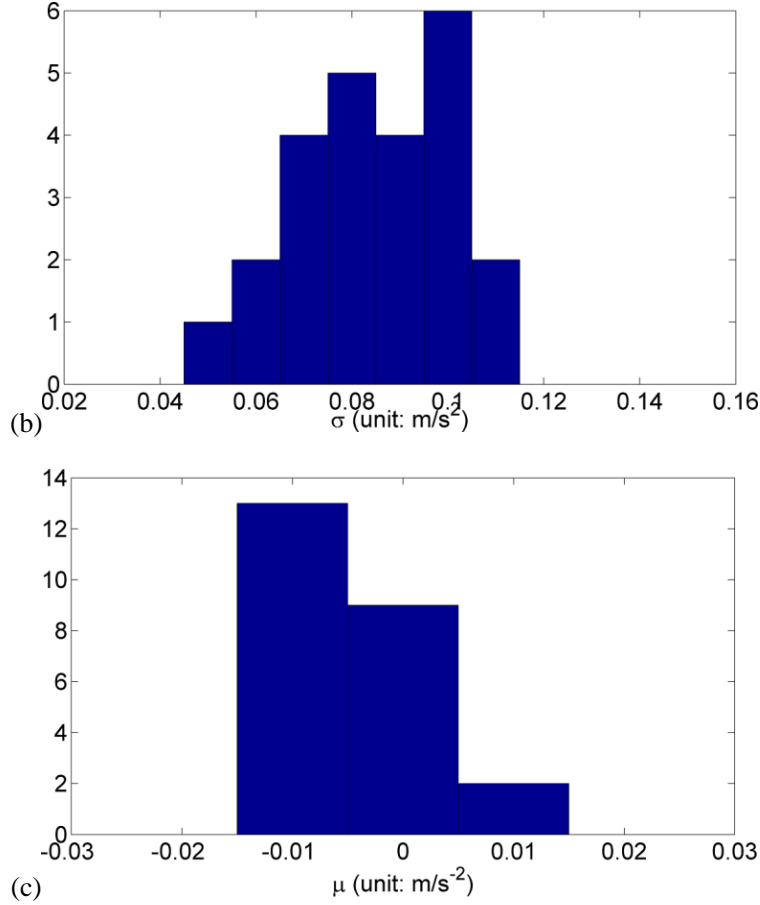

**Figure 8.** Statistical results of calibrated parameter (a) $\kappa$, (b) $\sigma$, (c) $\mu$.

## 3. The explanation of concave growth of the speed variance in the platoon

Using our proposed model and some additional assumptions, we explain here why the growth of the speed variance observed in our car-following experiments show a concave pattern. The assumptions are (i) there exists an indifference region, in which drivers have no response to the changes in spacings; (ii) a driver adapts his/her vehicle's speed to that of the preceding vehicle ; (iii) a car's new speed is also subject to independent stochastic contributions leading to an additional speed stochasticity $\xi$ with mean $<\xi>=0$ and variance $Var(\xi)=\sigma^2$. In assumption (ii), the reaction delay has not been considered, since the reaction delay only has slight quantitative influence on the oscillation growth pattern, see the simulation section below.

We assume a platoon consisting of $N$ vehicles following a leader driving at a constant speed $v$, where vehicle 1 is the first follower and vehicle $N$ is the end of the platoon. The speed of vehicle $n$ is $v_n$. According to the above assumptions, the fluctuating speed of vehicle 1 is given by $v_1 + \xi_1 = v$, where $\xi_1$ represents the stochasticity of this vehicle. Furthermore, vehicle 2 will move at the speed $v_2 + \xi_2 = v_1$, and vehicle $n$ will move at the speed $v_n + \xi_n = v_{n-1}$, i.e.,

$$v_n = v - \xi_1 - \ldots - \xi_n \tag{10}$$

Since all $\xi_i$ ($i = 1,..., n$) are independent, we obtain

$$<v_n> = v - n<\xi> = v \tag{11}$$

$$\text{Var}(v_n) = n\text{Var}(\xi) = n\sigma^2 \tag{12}$$

Thus, the standard deviation of the speed of vehicle $n$ increases in a concave way along the platoon.

$$\sigma_n = \sqrt{n}\sigma \tag{13}$$

## 4 The car following model

Based on above analysis, we propose a simple car following model as follows: (i) when the spacing between vehicles is in the indifference region [$d^{\min}$, $d^{\max}$], the acceleration of vehicles will be determined by the speed adaptation and the stochastic part; (ii) when the spacing is outside the indifference region, the normal deterministic car following acceleration is used. Taking the revised IDM (Horiguchi and Oguchi, 2014) as an example, the new model is specified as follows:

$$\begin{aligned}
&\text{if } d_n^{\min} \leq d_n \leq d_n^{\max} \\
&\quad a_n = \min\left(\lambda \Delta v_n(t-\tau) + \xi_n, \lambda(v_{\max} - v_n)\right) \\
&\text{else} \\
&\quad a_n = \min\left(a\left(1 - \left(\frac{v_n}{v_{\max}}\right)^4\right), a\left(1 - \left(\frac{d_n^{\min} + d_n^{\max}}{2d_n}\right)^2\right)\right)
\end{aligned} \tag{14}$$

where $\xi_n$ is determined by Eq. (7). The boundaries of the indifference region are determined by

$$d_n^{\min} = \max\left(v_n(t)T_{\min} - \frac{v_n(t)\Delta v_n(t)}{2\sqrt{ab}}, 0\right) + s_0 \tag{15}$$

$$d_n^{\max} = \max\left(v_n(t)T_{\max} - \frac{v_n(t)\Delta v_n(t)}{2\sqrt{ab}}, 0\right) + s_0 \tag{16}$$

In the indifference region, the term $\lambda(v_{\max} - v_n)$ is applied to ensure that the vehicle's speed $v_n$ will not exceed the maximum speed $v_{\max}$.

**Table 1.** model parameter values.

| Parameter | Description | Unit | Value |
|---|---|---|---|
| $v_{max}$ | maximum speed | m/s | 30 |
| $T_{min}$ | minimum desire time gap | s | 0.6 |
| $T_{max}$ | maximum desire time gap | s | 1.7 |
| $a$ | maximum acceleration | m/s$^2$ | 0.85 |
| $b$ | desire deceleration | m/s$^2$ | 1.5 |
| $s_0$ | minimum standing gap | m | 2.0 |
| $\tau$ | reaction delay | s | 1.3 |
| $\lambda$ | sensitivity | s$^{-1}$ | 0.27 |
| $\kappa$ | sensitivity | s$^{-1}$ | 0.09 |
| $\sigma$ | acceleration strength | m/s$^2$ | 0.07 |

## 5. Simulation results

### 5.1 oscillations in car following platoon

To quantitatively reproduce the concave growth of speed variance or standard deviations (STDs) in the platoon, we calibrate the model parameters, using the Generic algorithm. The fitness function is to minimize the difference between the simulated STDs and experimental ones. The calibrated parameters are shown in Table 1. Moreover, the Fast Fourier transform analysis (FFT, Li, et al. 2010) of the speed time series of the last vehicle was performed to compare the simulated and experimental oscillation frequencies in the platoon. Fig. 9 and 10 show that the concave growth pattern is quantitatively reproduced by the new model, and the spatiotemporal evolution of traffic flow has been captured.

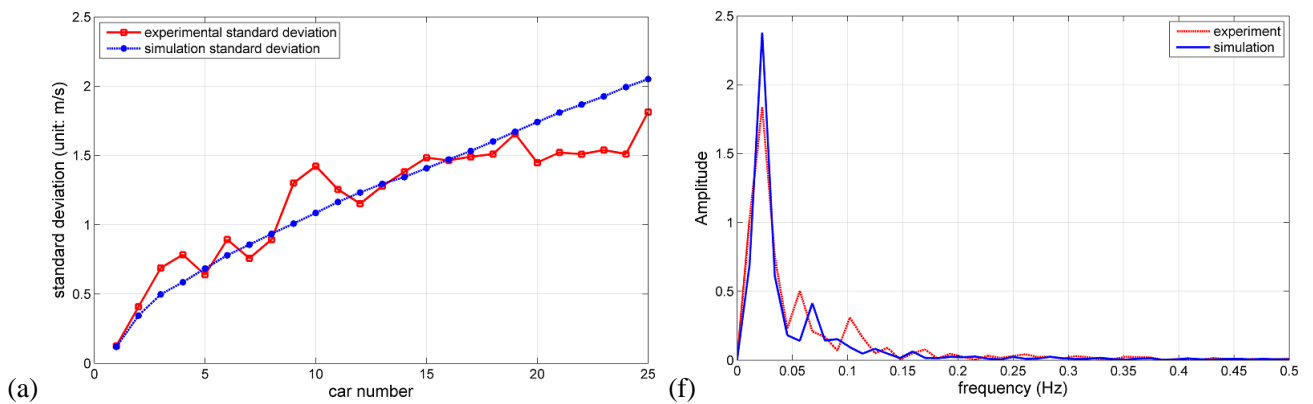

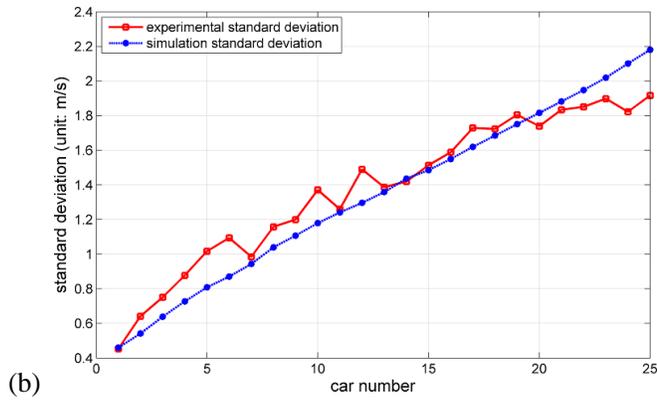
(b)

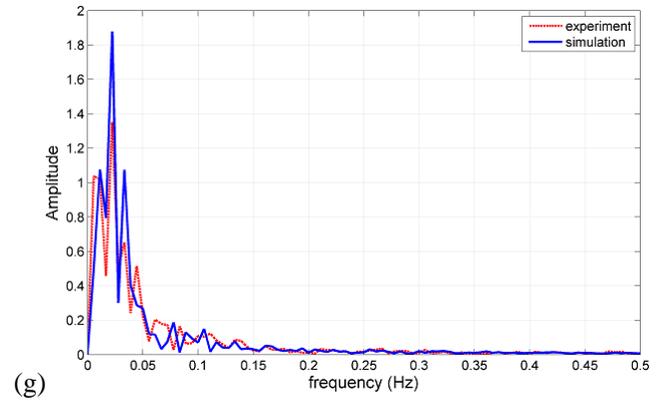
(g)

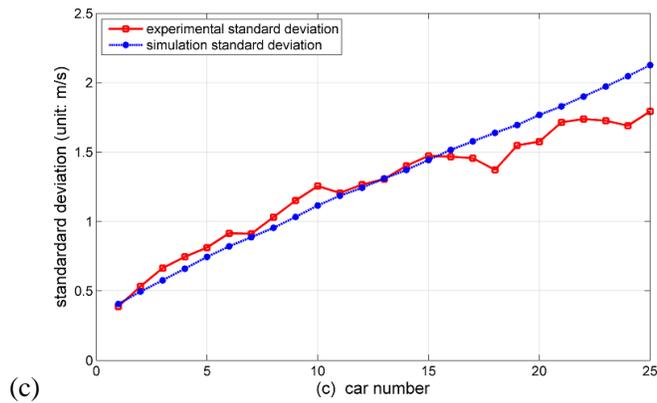
(c)

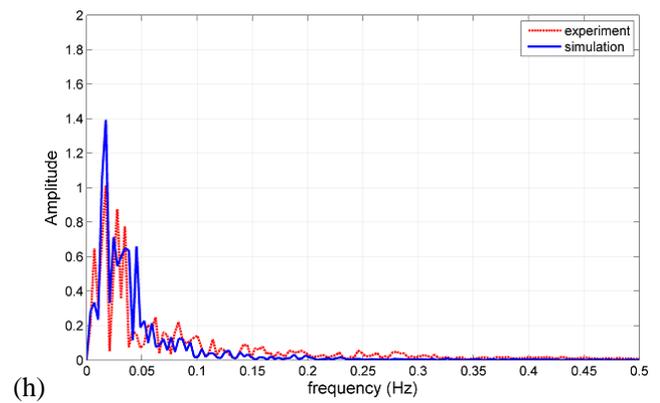
(h)

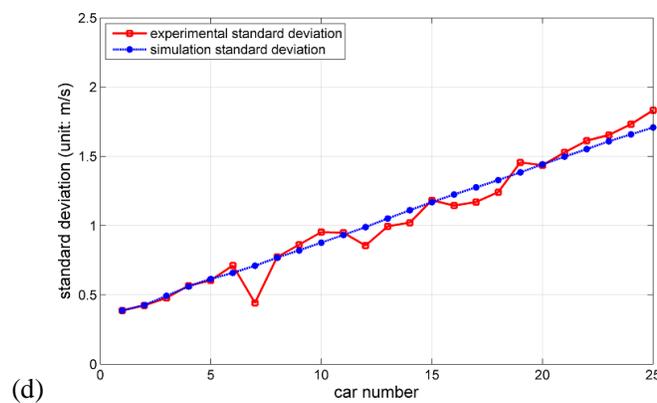
(d)

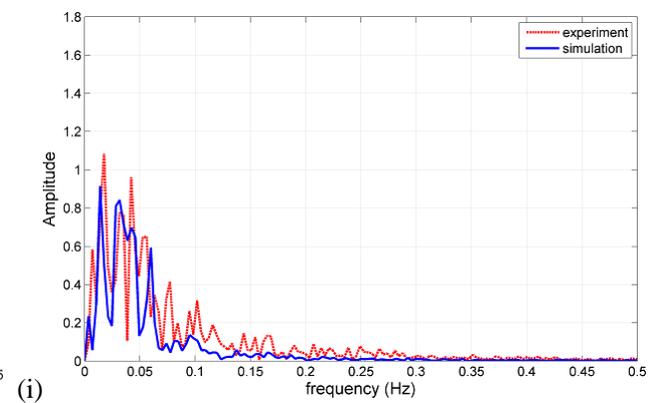
(i)

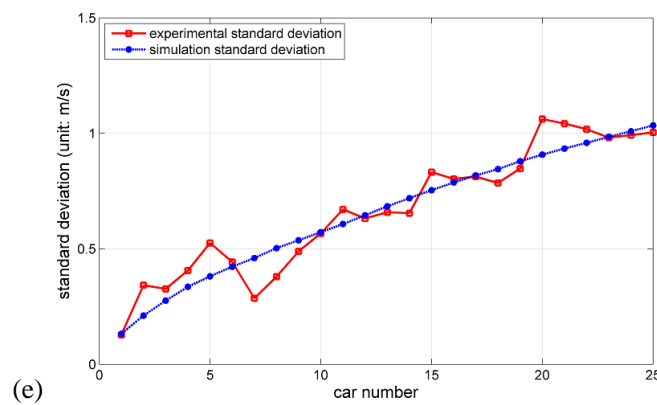
(e)

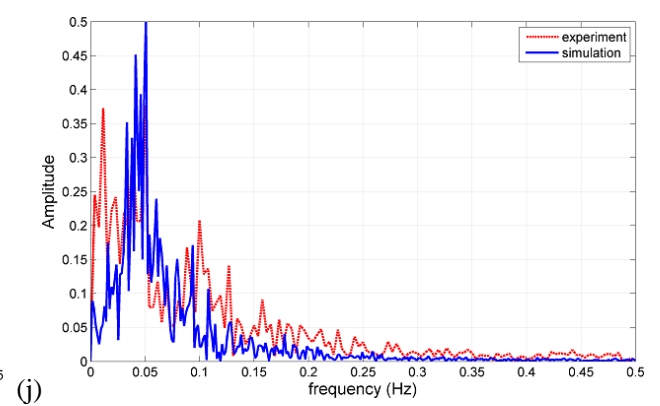
(j)
15

**Figure 9.** (a)-(e) Standard deviations of the car following platoons; (f)-(i) the corresponding FFT spectra for detrended speed data of the last vehicle. The car number 1 is the leading car. From (a) to (e) and (f) to (j), the leading car moves with $v_{leading}$=50, 40, 30, 15 and 7 km/h, respectively.

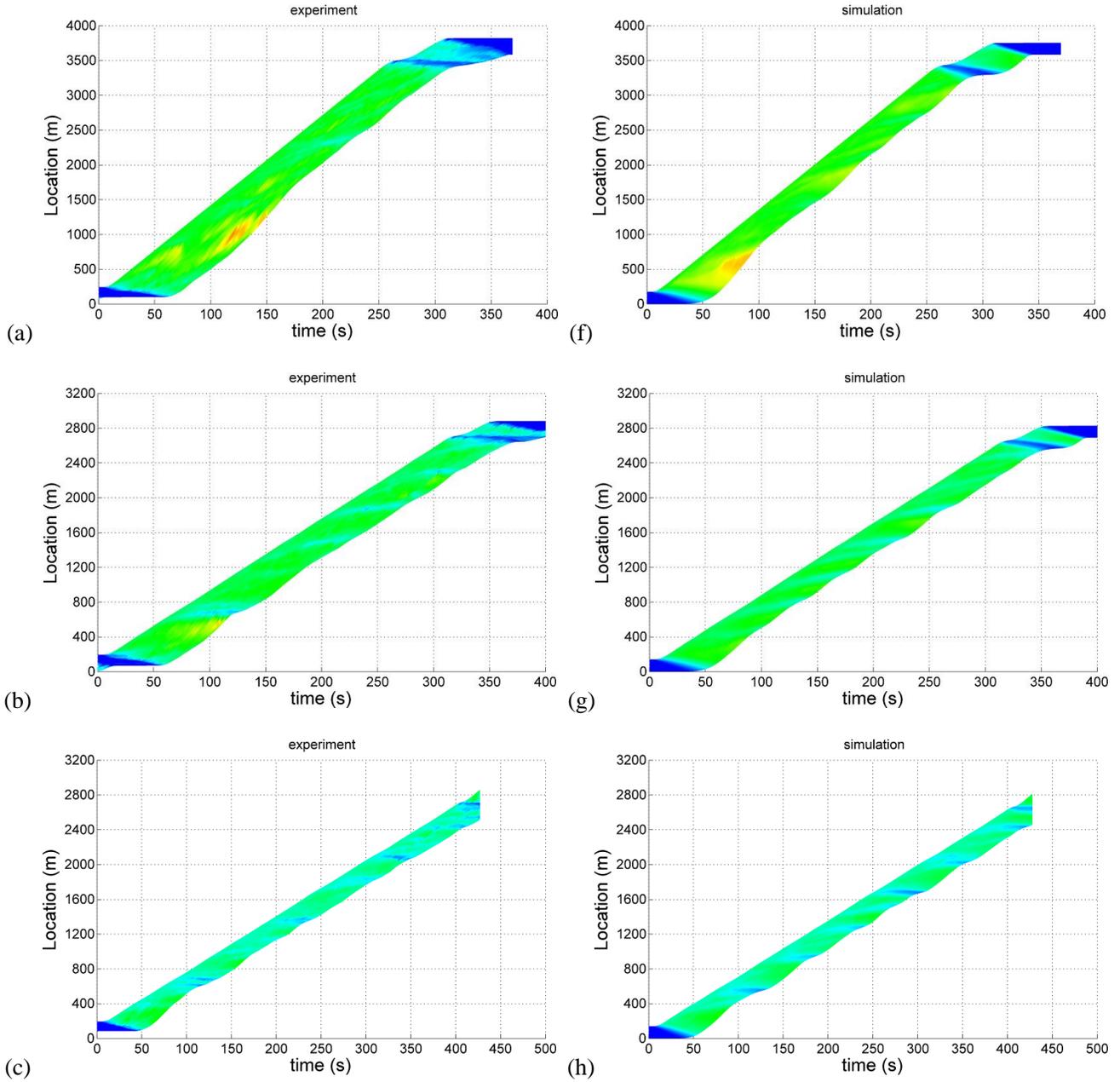

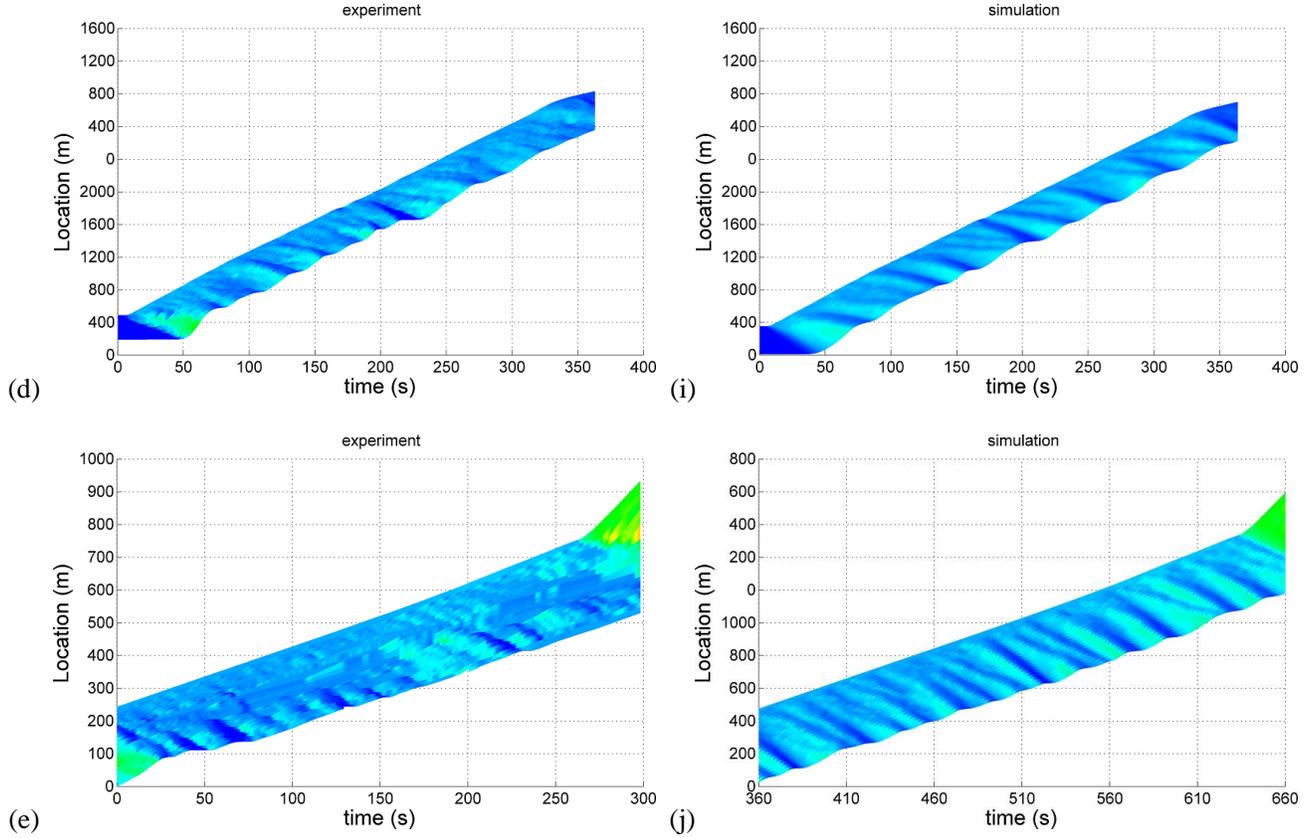

**Figure 10.** The spatiotemporal diagrams of the car following platoons. (a)-(e) Experimental results; (f)-(i) simulation ones. From (a) to (e) and (f) to (j), the leading car moves with $v_{\text{leading}}$=50, 40, 30, 15 and 7 km/h, respectively.

*5.2 Discharge Rate*

Yuan et al. (2015, 2017) reported a linear relationship between the speed in congestion and the queue discharge rate: the queue discharge rate increases with the speed in congestion. Next we examine whether this feature can be captured in the new model or not.

In the simulation, a 25-car following platoon has been simulated. The leading car moves with speed $v_{\text{leading}}$ when its location $x < x_0$. As a result, average speed of the congested traffic in the platoon equals $v_{\text{leading}}$. The leading car accelerates freely when $x > x_0$. The discharge rate is calculated by $24/(t_2-t_1)$, where $t_1$ and $t_2$ denote the time for the leading car and the last car to pass the location $x = x_{\text{de}}$ downstream of $x_0$, respectively, see Fig. 15(a). Fig. 15(b) shows that the linear relationship between the congested speed and discharge rate is well simulated.

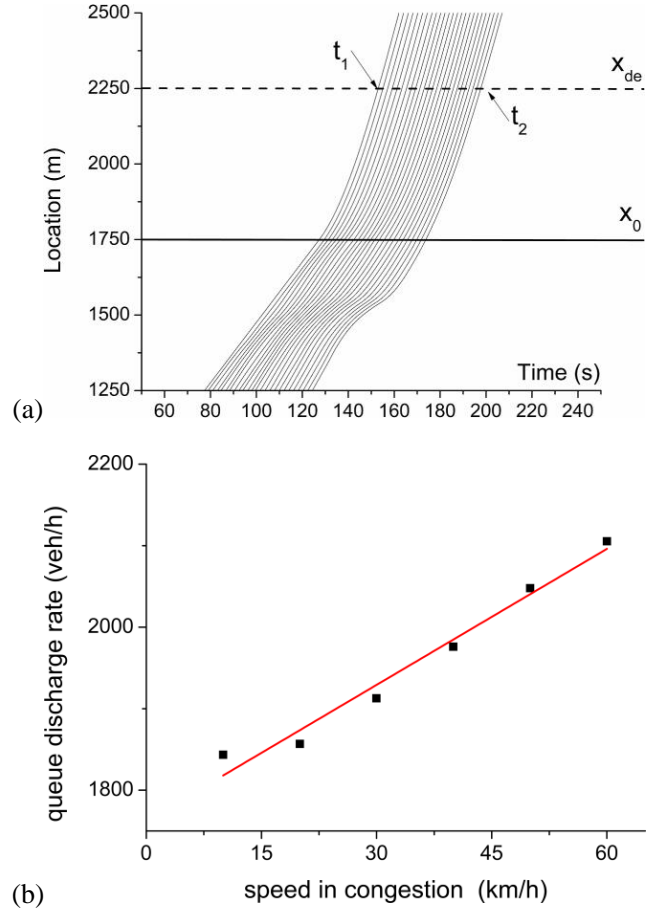

**Figure 11.** (a) An sample of the trajectories of the queue discharge simulation. (b) the relation between queue discharge rate and the speed of congested traffic. The two locations $x_0 = 1750$ m, $x_{de} = 2250$ m. In the sample in (a), $v_{leading} = 10$ m/s.

*5.3 Sensitivity analysis*

This subsection performs a sensitivity analysis to examine the role of reaction delay and indifference region boundary in traffic flow dynamics.

Firstly, their influence on the oscillation growth is investigated. Fig. 12 shows the impact of reaction delay in the case $v_{leading}$=50 km/h. One can see that reaction delay only has slight effect. The oscillation only grows a little faster even if reaction delay increases from 0.5 s to 2.0 s. In the case of other leading speed, similar results were observed.

Fig. 13 shows the influence of the indifference region boundary. The lower boundary depends on $T_{min}$ and the upper one depends on $T_{max}$. The boundary has significant impact on the oscillation growth rate. With the increase of $T_{max}$ or the decrease of $T_{min}$ (i.e., with the expansion of indifference region), the oscillation grows much faster.

Next, we study the impacts of reaction delay and indifference region boundary on the discharge rate. Fig. 14(a) and (b) shows that $\tau$ and $T_{min}$ has only slight quantitative impact on the discharge rate. However, the discharge rate decreases significantly with the increase of $T_{max}$, as shown in Fig.14(c).

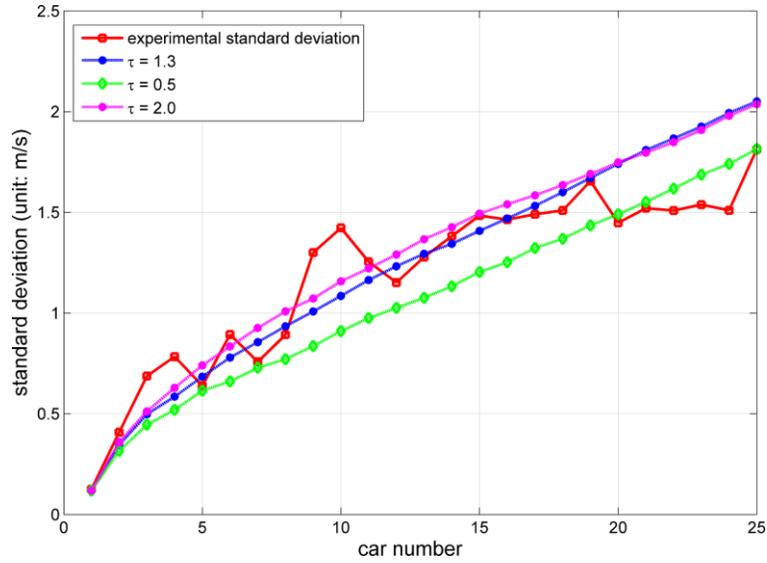

**Figure 12.** Standard deviations of the car following platoon. The platoon leader moves with $v_{leading}$=50 km/h.

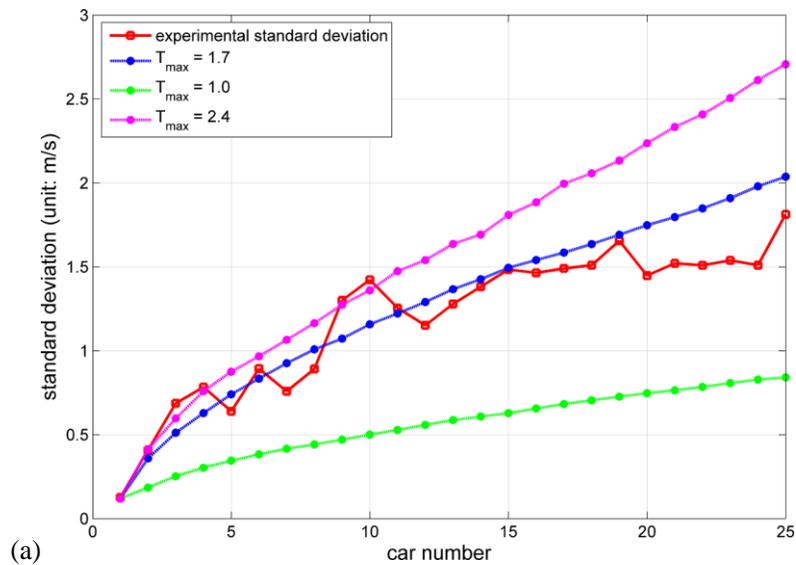

(a)

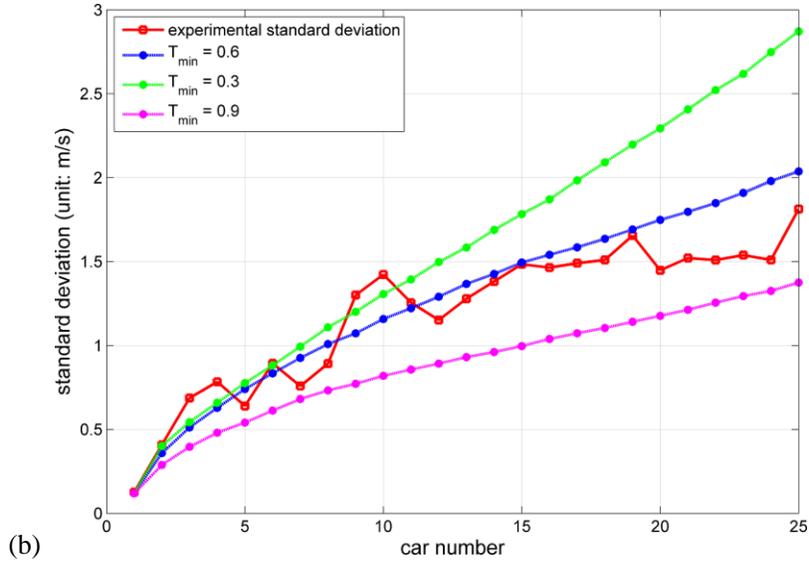

(b)

**Figure 13.** Standard deviations of the car following platoon. The platoon leader moves with $v_{\text{leading}}$=50 km/h. In the case of other leading speed, similar results were observed. (a) The impact of $T_{\max}$; (b) The impact of $T_{\min}$.

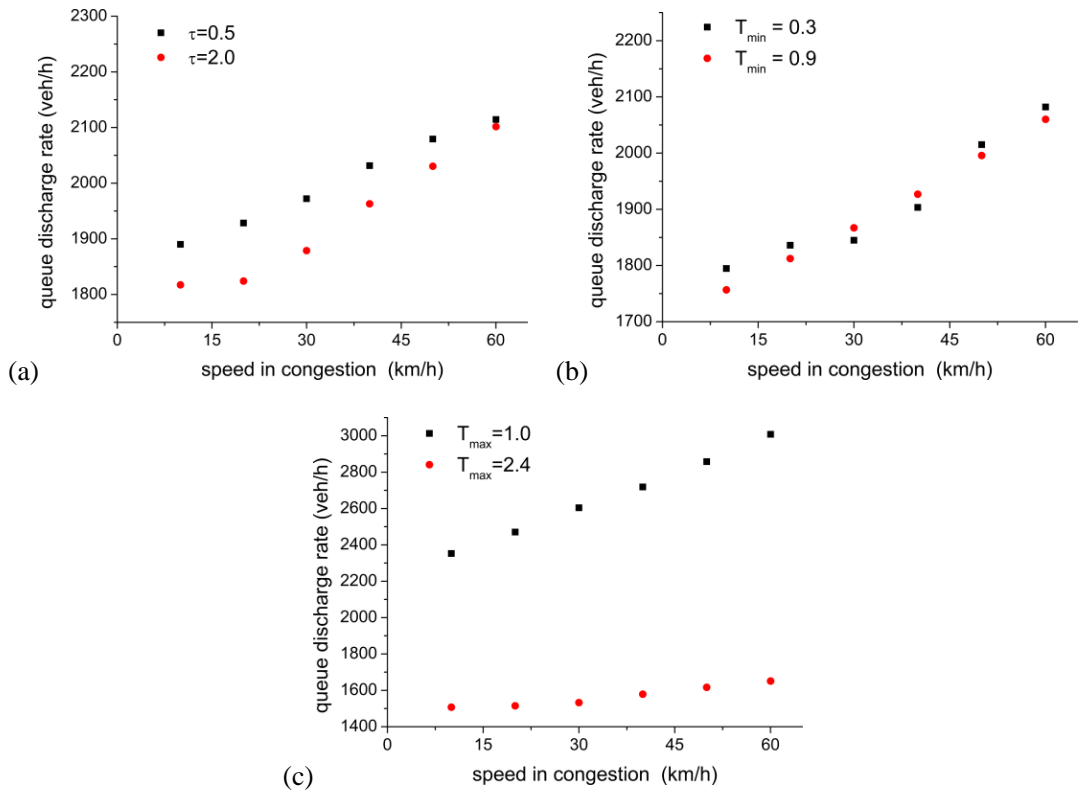

**Figure 14.** Impact of (a) $\tau$, (b) $T_{min}$, (c) $T_{max}$ on discharge flow rate.

## 6. Conclusion

In previous works, we have proposed a new instability mechanism based on our car-following experiments, i.e., traffic instability is determined by the speed adaptation effect and the cumulative effect of stochastic factors. In this paper, we provided a quantitative analysis of this mechanism based on a new stochastic car-following model supported by a thorough analysis of the trajectory data in the car-following experiment.

Firstly, we compare the time series of acceleration with the speed difference. It is shown that the two time series exhibit striking similarity, with acceleration exhibiting a time delay. The time delay is obtained using correlation and found to follow the lognormal distribution. We find that the speed difference plays a more important role on car-following dynamics than the spacing. The sensitivity parameter $\lambda$ is calculated through linear regression of $a=\lambda \Delta v(t-\tau)$, which also follows the lognormal distribution.

Secondly, the acceleration residual $\xi(t) = a - \lambda \Delta v(t-\tau)$ is analyzed. Results show that $\xi(t)$ can be regarded as a mean reversion process. To measure the strength of $\xi(t)$, the ratio of $\sigma_\xi/\sigma_a$ is calculated, which is quite large for the several vehicles in the front part of the platoon and decreases along the platoon. It indicates that when its amplitude is small, the growth of oscillations is mainly determined by the stochastic factors; When its amplitude increases, the growth of the oscillations is determined by the competition between the stochastic factors and the speed difference

Based on these findings, we provided an explanation for why the speed variance of vehicles in the platoon grow in a concave way along the platoon, and proposed a mode-switching car-following model with two dominant modes of following behavior. Moreover, sensitivity analysis with respect to the reaction delay and the indifference region boundary has been performed. It was shown that reaction delay only has slight effect but indifference region boundary has significant on oscillation growth rate and discharge rate.

In the experiment, the oscillations are intrinsically generated, i.e., no external disturbance is exerted on the leading car. Therefore, it would be interesting and important to study the evolution of external oscillations in the future work. Moreover, it is also necessary to find out boundary of the indifference region.

**Acknowledgements:**

JFT was supported by the National Natural Science Foundation of China (Grant No. 71771168). RJ was supported by the Natural Science Foundation of China (Grant Nos. 71621001, 71631002).